\documentclass{main}

\usepackage[T1]{fontenc}
\usepackage{lineno}
\modulolinenumbers[5]
\usepackage{amsmath, amssymb}
\usepackage{xspace}
\usepackage{csquotes}

\usepackage[sorting=none]{biblatex}
\AtBeginBibliography{\small}
\DeclareFieldFormat[online]{title}{\mkbibquote{#1}}

\def\Jpsi{\ensuremath{\text{J}/\psi}\xspace}
\def\pT{\ensuremath{p_{\rm{T}}}\xspace}
\def\mant{\ensuremath{\text{|}t\text{|}}\xspace}

\newcommand{\Photo}{}

\begin{document}

\title{\Large Recent results on ultra-peripheral collisions with the ALICE experiment}

\author{David Grund (for the ALICE Collaboration)}

\address{Faculty of Nuclear Sciences and Physical Engineering, Czech Technical University in Prague \\ Prague, Czech Republic}

\maketitle\abstracts{Recent measurements of \Jpsi photoproduction based on data from ultra-peripheral Pb--Pb and p--Pb collisions recorded by the ALICE experiment during Run~2 of the LHC are presented. Photoproduction as a photon-induced process is sensitive to the structure of hadrons and the results are of great importance for a better understanding of how gluon saturation and nuclear shadowing work in high-energy quantum chromodynamics. The energy evolution of coherent \Jpsi photoproduction has been measured, indicating a strong suppression of nuclear gluon densities at high energies. The energy dependence has also been measured for exclusive and dissociative \Jpsi production off the proton. The average distribution of the nuclear gluon matter in transverse plane, and fluctuations thereof, have been studied for the first time using the measurements of the dependence of coherent and incoherent \Jpsi photoproduction on the transferred momentum \mant. Lastly, the data from coherent photoproduction of \Jpsi in Pb--Pb ultra-peripheral collisions have been found to be compatible with a transverse polarization.}


\section{Introduction}
When the hadronic structure is studied using a high-energy probe, corresponding to low values of Bjorken-$x$ of the colliding parton, the gluon density becomes so large that the regime of gluon saturation~\cite{Morreale:2021pnn}---a dynamic equilibrium between gluon splitting and recombination---is predicted to be reached. Additionally, when nuclei at high energies are considered, the parton distribution functions of constituent nucleons are modified with respect to free nucleons by the presence of the nuclear environment; this phenomenon is referred to as nuclear shadowing~\cite{Armesto:2006ph}. Ultra-peripheral collisions (UPCs) allow to shed light on these high-energy quantum chromodynamics (QCD) effects using photon-induced processes with large cross sections that are sensitive to the gluon distribution in the target proton or nucleus.

In UPCs, the nuclei collide at an impact parameter $b$ larger than the sum of their radii, $b > R_1 + R_2$, so that short-range hadronic interactions are suppressed. The interactions are thus induced by quasireal photons (with low virtuality) emitted by one of the nuclei. The intensity of the photon flux scales with the nuclear charge squared, $Z^2$, while the photon energy increases with the beam energy (the Lorentz boost). Lead nuclei collided at the LHC thus deliver very strong electromagnetic fields. 

Diffractive production of vector mesons is one of the prominent UPC processes. The \Jpsi is an ideal candidate since it can be reconstructed with high precision using detectors of the ALICE experiment and its mass allows for application of perturbative QCD. If one focuses on the decay of \Jpsi into a pair of electrons or muons, such photoproduction events have a very clear experimental signature -- only two lepton tracks are present in an otherwise empty detector, except for potential products of electromagnetically dissociated nuclei in very forward rapidity regions.

The value of the momentum transferred in the interaction, the Mandelstam variable \mant, provides further insight into aspects of the photon-hadron interaction and can be used to classify photoproduction events. It approximately corresponds to the square of the transverse momentum of the vector meson, $\mant \approx \pT^2$; the relation is not exact due to the non-zero photon momentum in the laboratory frame. In studies involving high-\mant events, the photon contribution can usually be neglected. For Pb--Pb UPCs, the process is called either coherent if the photon couples coherently to all nucleons, which leads to $\mant \lesssim 0.01$~GeV$^2$, or incoherent if the photon couples to a single nucleon. In the latter case, \mant is governed by the size of the nucleon and its characteristic value corresponds to 0.1~GeV$^2$. The incoherent interaction leaves the target nucleus excited and is, in most cases, followed by a breakup. If the photon scatters from smaller---subnucleon-sized---objects, which leads to nucleon dissociation, \mant can easily reach the order of 1~GeV$^2$. In asymmetric p--Pb UPCs, the Pb nucleus is in most cases the photon emitter considering the $Z^2$ factor in the photon flux. If the photon interacts with the proton elastically, the average value of \mant amounts to 0.1~GeV$^2$; values of the order of 1~GeV$^2$ are reached in the case of proton dissociation.

\section{Ultra-peripheral collisions at ALICE}
The ALICE experiment~\cite{ALICE:2008ngc} at the LHC is well equipped to measure UPC processes. The detectors crucial for UPC studies using data from Run 2~(2015--2018) of the LHC are briefly described here. The central-barrel detectors, placed inside a large solenoid magnet creating a magnetic field of 0.5~T, cover together the central pseudorapidity region of $|\eta| < 0.9$. The central-barrel detectors detailed here have a cylindrical geometry and cover the full azimuth. The Silicon Pixel Detector (SPD) constitutes the two innermost layers of the Inner Tracking System (ITS) and is used for tracking, vertex reconstruction, and provides trigger inputs. Its inmost layer is located at a radius of only 3.9~cm from the beam axis. Reconstruction of tracks further away from the IP and their identification by measuring ionization energy losses is provided by the Time Projection Chamber (TPC), a large cylindrical gas chamber equipped with multi-wire proportional chambers mounted on end caps. The Time-Of-Flight (TOF) detector consisting of multi-gap resistive-plate chambers is placed outside the TPC. TOF is a fast detector and provides topological trigger signals in UPC measurements based on the number of azimuthal sectors registering a signal.

ALICE also uses a single-arm spectrometer detecting forward muons in the $-4 < \eta < -2.5$ pseudorapidity region. It is composed of an absorber with a thickness of 10 hadronic interaction lengths, followed by five tracking stations, each consisting of two planes of cathode-pad chambers. The middle station is housed in a dipole magnet providing an integrated magnetic field of 3~T$\cdot$m. The muon trigger system is located at the rear of the spectrometer, separated by another muon filter in the form of an iron wall with a thickness of 7.2 hadronic interaction lengths.

The trigger configurations used for UPCs rely heavily on inputs from two forward systems of scintillation counters, V0 and ALICE Diffractive (AD). These detectors have two components (A and C) located near the beam pipe, one on each side of the IP, and operate as a veto. No activity above a certain threshold is thus allowed at the time of the collision to ensure that UPC processes are not contaminated by hadronic interactions or inelastic photonuclear processes. Additionally, the Zero Degree Calorimeters (ZDCs) installed 113~m away from the IP along the beam in both directions are used to detect protons and neutrons in very forward regions emitted from nuclear breakups.

\section{Energy dependence of \texorpdfstring{\Jpsi}{J/psi} photoproduction}

Photoproduction of \Jpsi in Pb--Pb UPCs is sensitive to the nuclear gluon density at Bjorken-$x$ of $x = M_{\Jpsi}^2/W_{\gamma{\rm Pb,n}}^2$, where $M_{\Jpsi}$ is the \Jpsi mass. The photon-nucleus center-of-mass energy per nucleon pair, $W_{\gamma{\rm Pb,n}}$, is related to the \Jpsi rapidity $y$ through
\begin{equation}
    W_{\gamma{\rm Pb,n}}^2 = \sqrt{s_{\rm NN}} M_{\Jpsi} e^{-y} \, ,
\end{equation}
where $\sqrt{s_{\rm NN}}$ is the center-of-mass energy of the collision system per nucleon pair. The rapidity is given in the laboratory frame and is defined with respect to the direction of the target. Since it is not known which ion emitted the photon and which was the target, a twofold ambiguity arises. The \Jpsi produced at a rapidity $y$ thus probes the gluons at two values of $x$, 
\begin{equation}
x_{1,2} = \frac{M_{\Jpsi}}{\sqrt{s_{\rm NN}}} e^{\pm |y|} \, .
\label{eq:Bjx}
\end{equation}
The cross section for \Jpsi photoproduction also includes both contributions, 
\begin{equation}
    \frac{\rm{d}\sigma_{\rm PbPb}}{{\rm d} y} = n_{\gamma}(y, \{b\}) \sigma_{\gamma{\rm Pb}}(y) + n_{\gamma}(-y, \{b\}) \sigma_{\gamma{\rm Pb}}(-y) \, ,
\label{eq:upc_cs}
\end{equation}
where $n_{\gamma}$ is the photon flux, $\sigma_{\gamma\rm{Pb}}$ is the photonuclear cross section and $\{b\}$ denotes the impact parameter range taken into account in the measurement. 

The ambiguity in the photon energy needs to be addressed in order to measure the energy (or $x$) dependence of \Jpsi photoproduction. The issue is not present in studies of Pb--Pb UPCs at $y \approx 0$ since the two contributions can be effectively added up. The same applies to p--Pb collisions, where the term describing the photon emission from the proton can be neglected. For analyses of Pb--Pb UPC data in forward rapidity regions, the general strategy is to combine several measurements at the same rapidity but in different impact-parameter ranges, provided that the corresponding photon fluxes are calculated. This allows to determine the values of the photonuclear cross section $\sigma_{\gamma{\rm Pb}}(y)$ and $\sigma_{\gamma{\rm Pb}}(-y)$ upon solving a system of linear equations, given by Eq.~\eqref{eq:upc_cs} for the respective impact-parameter ranges.

Two possible implementations of this approach are based either on a combination of results from UPC and peripheral collisions ($b < R_1 + R_2$, with a partial nuclear overlap)~\cite{Contreras:2016pkc} or on tagging of events using forward neutrons~\cite{Guzey:2013jaa}. The latter method makes use of the fact that the electromagnetic fields of the colliding nuclei are so strong that an independent photon exchange can lead to an electromagnetic dissociation (EMD) of at least one of the nuclei, which can be accompanied by the emission of forward neutrons registered in the ZDCs. Events are then classified as 0n0n (no neutrons on either side), 0nXn or Xn0n (neutrons on one side only) or XnXn (forward neutrons on both sides). The photon spectra corresponding to these fragmentation scenarios can be calculated~\cite{Baltz:2002pp} and occupy different impact-parameter ranges.

\subsection{Coherent \texorpdfstring{\Jpsi}{J/psi} production}
Recently, ALICE has measured the energy dependence of coherent \Jpsi photonuclear production based on simultaneous analysis of Run-2 data from Pb--Pb UPCs at $\sqrt{s_{\rm NN}} = 5.02$~TeV at central and forward rapidities~\cite{ALICE:2023jgu}. Using the separation into neutron classes, the dependence of the cross section on $W_{\gamma{\rm Pb,n}}$, the center-of-mass energy of the $\gamma{\rm Pb}$ system per nucleon pair, was determined in the unprecedented interval of 17 to 920~GeV, corresponding to the $x$ range spanning three orders of magnitude, from $10^{-2}$ down to $10^{-5}$. The CMS Collaboration has performed a similar measurement in a narrower energy range~\cite{CMS:2023snh}.

\begin{figure}
\centering
\includegraphics[width=0.49\textwidth]{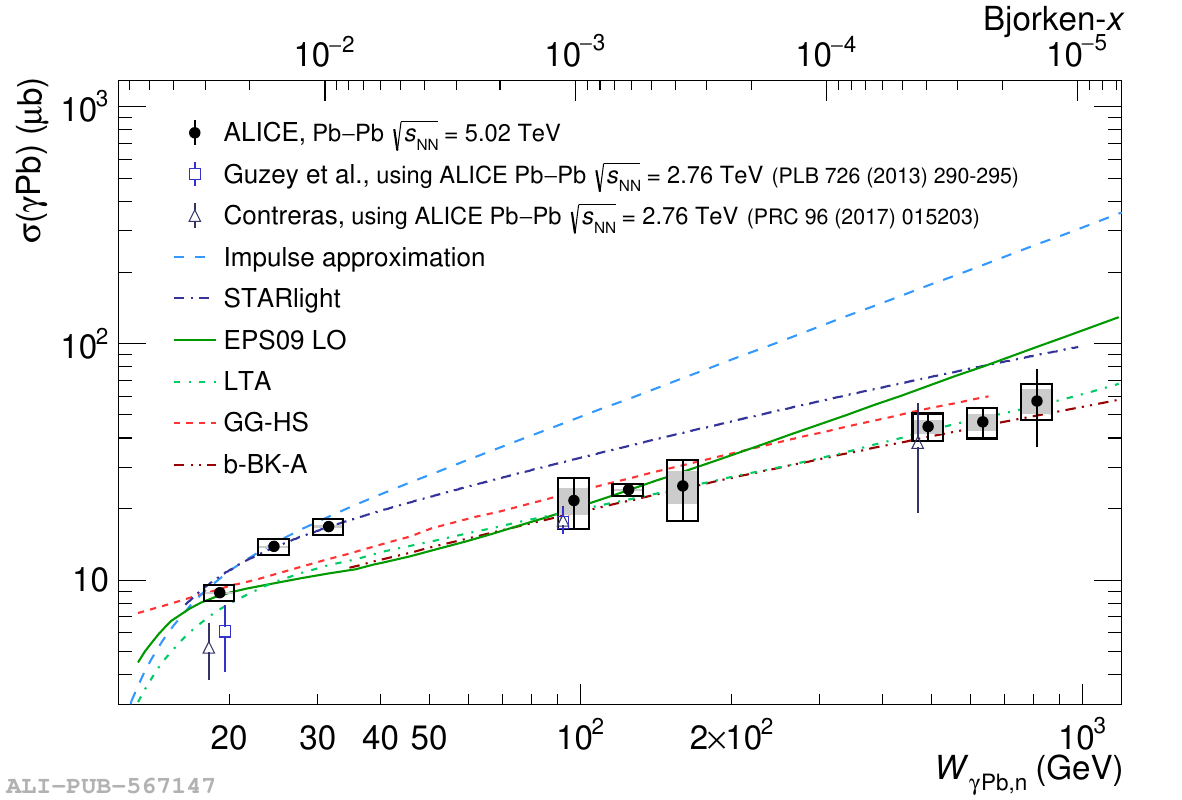}
\includegraphics[width=0.49\textwidth]{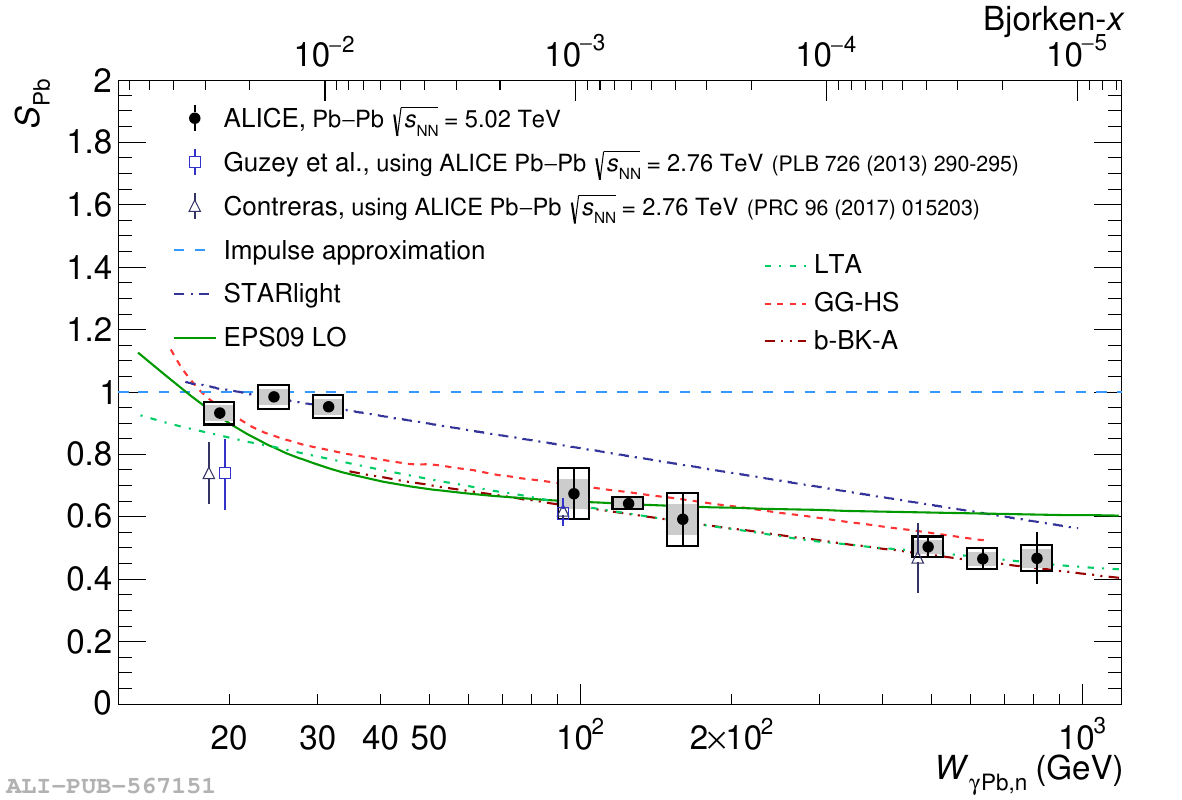}
\caption{The dependence of the cross section for coherent \Jpsi photoproduction on photon-nucleus energy per nucleon pair, $W_{\gamma{\rm Pb,n}}$, or $x$ as measured by ALICE in Pb--Pb UPCs at $\sqrt{s_{\rm NN}} = 5.02$~TeV~\protect\cite{ALICE:2023jgu} (left). The measured energy and $x$ dependence of the nuclear suppression factor $S_{\rm Pb}$~\protect\cite{ALICE:2023jgu} (right). The vertical lines represent uncorrelated uncertainties, the empty boxes show the quadratic sum of the correlated systematic uncertainties and those coming from migration across neutron classes. The gray boxes show the theoretical uncertainty arising from the photon flux calculation. See the text for the description of the models.}
\label{fig:coh_sPb}
\end{figure}

The measured dependence is shown in the left panel of Figure~\ref{fig:coh_sPb} and is compared with predictions of various models. The Impulse approximation (IA)~\cite{Chew:1952fca} describes well the low-energy data but significantly overshoots all other points at higher energies, clearly marking the onset of nuclear shadowing. STARlight~\cite{Klein:2016yzr}, a hadronic model based on Glauber calculation, also describes properly only the low-energy points, implying the importance of including dynamic QCD effects as in the other models. The remaining models underestimate the cross section around 30 GeV, but provide a reasonably good description in other regions. The first model by the GSZ group~\cite{Guzey:2016piu} is based on EPS09 LO parametrization of nuclear PDFs, the latter incorporates leading twist approximation (LTA) of gluon shadowing. The GG-hs model~\cite{Cepila:2017nef} relies on a color-dipole approach and includes gluon saturation; the nucleons are pictured as a sum of hot spots, positions of which fluctuate event-by-event and their number increases with energy. The last model, b-BK-A~\cite{Bendova:2020hbb}, is based on a solution to the impact-parameter dependent Balitsky-Kovchegov (BK) equation and is valid only at $x$ smaller than $10^{-2}$. The new data are also compatible with Run-1 ALICE results, which were obtained by combining data from UPCs and peripheral collisions. 

The results were used to calculate the nuclear suppression factor $S_{\rm Pb}$, which gives a quantitative estimate of nuclear shadowing. $S_{\rm Pb}$ can be calculated as a square root of the ratio between the measured photonuclear cross section and the prediction given by IA, which ignores nuclear-environment effects. The measured energy dependence of $S_{\rm Pb}$ is depicted in the right panel of Figure~\ref{fig:coh_sPb}; while $S_{\rm Pb}$ is approximately 0.95 around 20--40 GeV ($x$ of $10^{-2}$), it drops with increasing energy and levels off at about 0.5. Note that the value of 0.65 measured in the center point ($W_{\gamma{\rm Pb,n}} = 125$~GeV) is compatible with the results derived from the previous ALICE midrapidity study of coherent \Jpsi production~\cite{ALICE:2021gpt}. 

\subsection{Exclusive and dissociative \texorpdfstring{\Jpsi}{J/psi} production}
\begin{figure}
\centering
\includegraphics[width=0.545\textwidth]{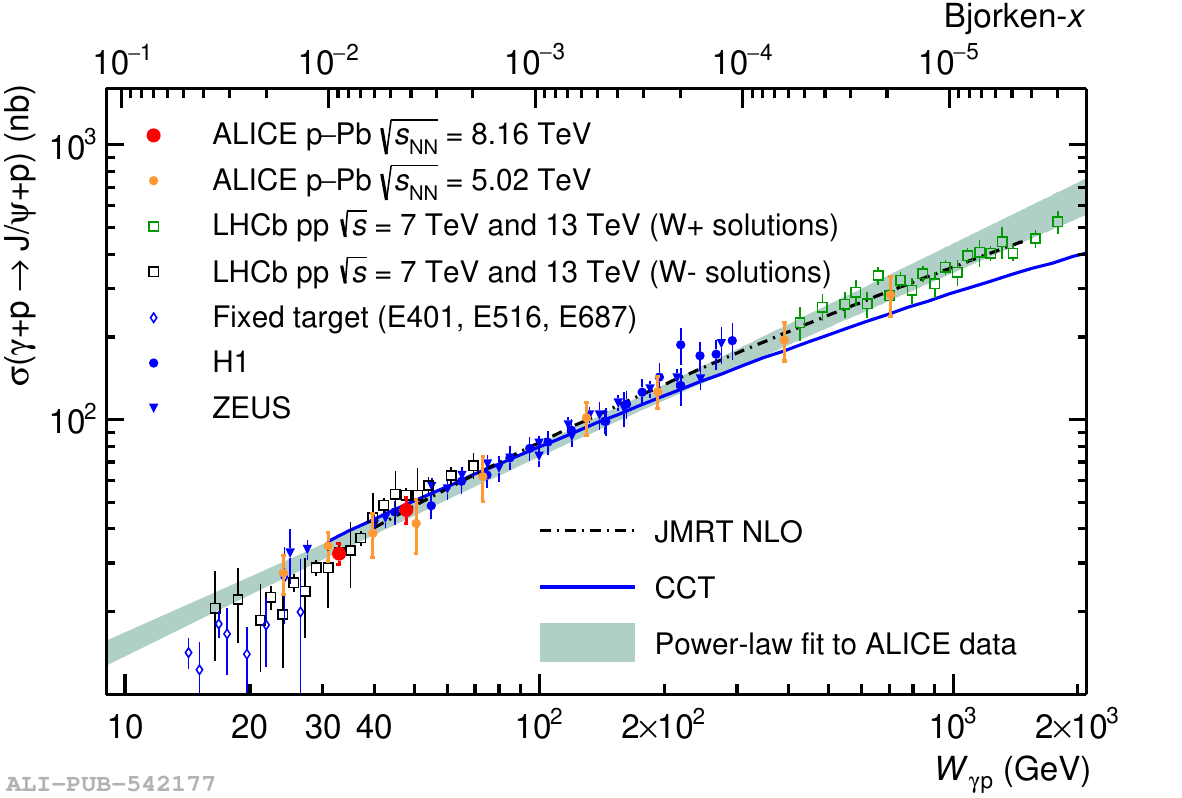}
\includegraphics[width=0.435\textwidth]{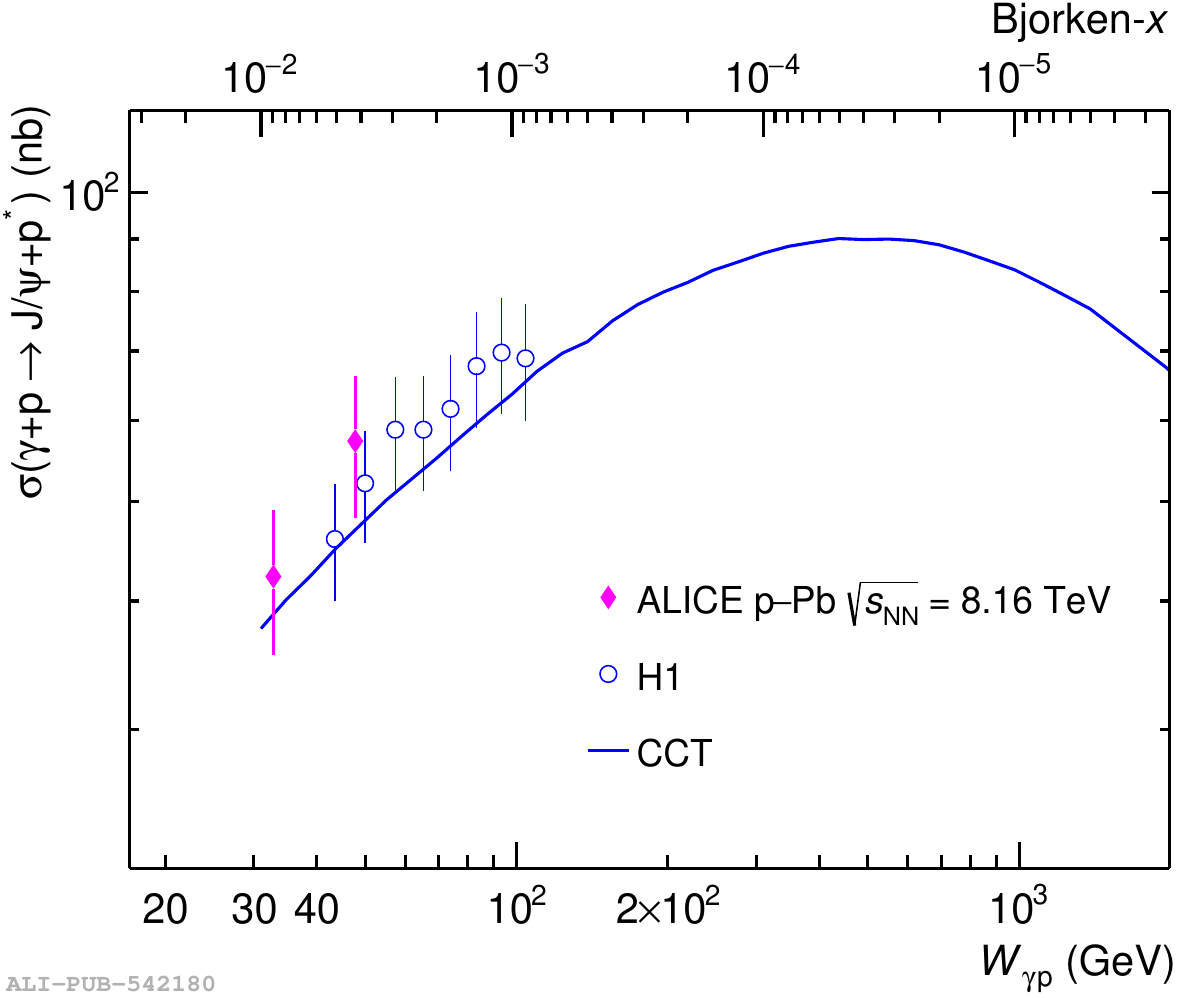}
\caption{The energy dependence of exclusive (left) and dissociative (right) \Jpsi photoproduction measured by ALICE in p--Pb UPCs at $\sqrt{s_{\rm NN}} = 8.16$~TeV~\protect\cite{ALICE:2023mfc}. The vertical lines represent the quadratic sum of the statistical and systematic uncertainties.}
\label{fig:excl_diss}
\end{figure}

The energy dependence of the cross section for \Jpsi photoproduction off protons has been studied by ALICE using p--Pb UPCs at $\sqrt{s_{\rm NN}} = 8.16$~TeV~\cite{ALICE:2023mfc}. Events with \Jpsi decaying into a pair of forward muons were registered with the muon spectrometer. The beam configuration corresponded to a low-energy photon emitted from the nucleus. Splitting the spectrometer acceptance into two rapidity bins, the cross section was determined at two values of $W_{\gamma{\rm p}}$, 27 and 57~GeV. The measured dependence of the exclusive \Jpsi cross section is shown in the left panel of Figure~\ref{fig:excl_diss}; the new data are compatible with the previous results from ALICE, LHCb, HERA and fixed-target experiments and, in addition, are consistent with a power-law behavior, $\sigma = N (W_{\gamma{\rm p}}/W_0)^\delta$, where $W_0 = 90$~GeV. The values of $\delta = 0.70 \pm 0.04$ and $N = 71.6 \pm 3.7$~nb were obtained by fitting the data.

The ALICE measurement of the energy dependence of \Jpsi photoproduction off protons accompanied by proton dissociation, shown in the right panel of Figure~\ref{fig:excl_diss}, is the first of its kind at a hadron collider. The observed energy evolution is in good agreement with previous H1 measurements and also with the CCT model~\cite{Cepila:2016uku}, where the proton substructure is again represented as a superposition of hot spots with fluctuating positions. The number of hot spots increases with the energy $W_{\gamma{\rm p}}$. Within the framework of the CCT model, the dissociative cross section is sensitive to fluctuations of subnucleon structures inside the proton~\cite{Miettinen:1978jb}. The model expects the cross section to reach a maximum at $W_{\gamma{\rm p}} \approx 500$~GeV, followed by a decrease. This is due to the hot spots effectively saturating the transverse area of the proton.

\section{Distribution of nuclear matter in the transverse plane}
Important insights into the structure of nuclei at high energies can be obtained by measuring the gluon distribution in the transverse plane. The dependence of the photonuclear cross section on Mandelstam \mant is related through a two-dimensional Fourier transform to the transverse-plane distribution of the matter in the target. In the Good-Walker approach~\cite{Good:1960ba}, coherent and incoherent photoproduction can be used to study the average and variance~\cite{Miettinen:1978jb} (quantum fluctuations) of the nuclear structure, respectively. Note that there may be caveats to the applicability of this approach, as discussed in Ref.~\cite{Klein:2023zlf}.
Since the value of \mant is inversely proportional to the size of the scattering center, gluon density fluctuations at the subfemtometer scale can be probed at $\mant \sim 1$~GeV$^2$. Should such fluctuations be significant, they would be reflected in an enhancement of the incoherent cross-section in the relevant region.

The ALICE Collaboration has recently performed the first measurement of the \mant-dependence of incoherent \Jpsi photoproduction in Pb--Pb UPCs at $\sqrt{s_{\rm NN}} = 5.02$~TeV~\cite{ALICE:2023gcs}, which complements the earlier ALICE study dedicated to the \mant-dependence of coherent \Jpsi production~\cite{ALICE:2021tyx}. Together, three orders of magnitude in \mant are covered by ALICE with an accuracy compatible with that at HERA. Both analyses are based on the same midrapidity data sample, sensitive to the $x$ interval of $(0.3, 1.4) \times 10^{-3}$. The analysis of coherent production focuses on the \Jpsi kinematic range of $\pT < 0.11$~GeV/$c$ and unfolding is employed to account for \pT migration and to correct for the photon transverse momentum. In the study of incoherent production, events with $0.2 < \pT < 1$~GeV/$c$ are selected, \pT resolution effects are negligible due to the widths of the considered intervals in \mant, and $\mant = \pT^2$ owing to large transferred momenta. 

\begin{figure}
\centering
\includegraphics[width=0.49\textwidth]{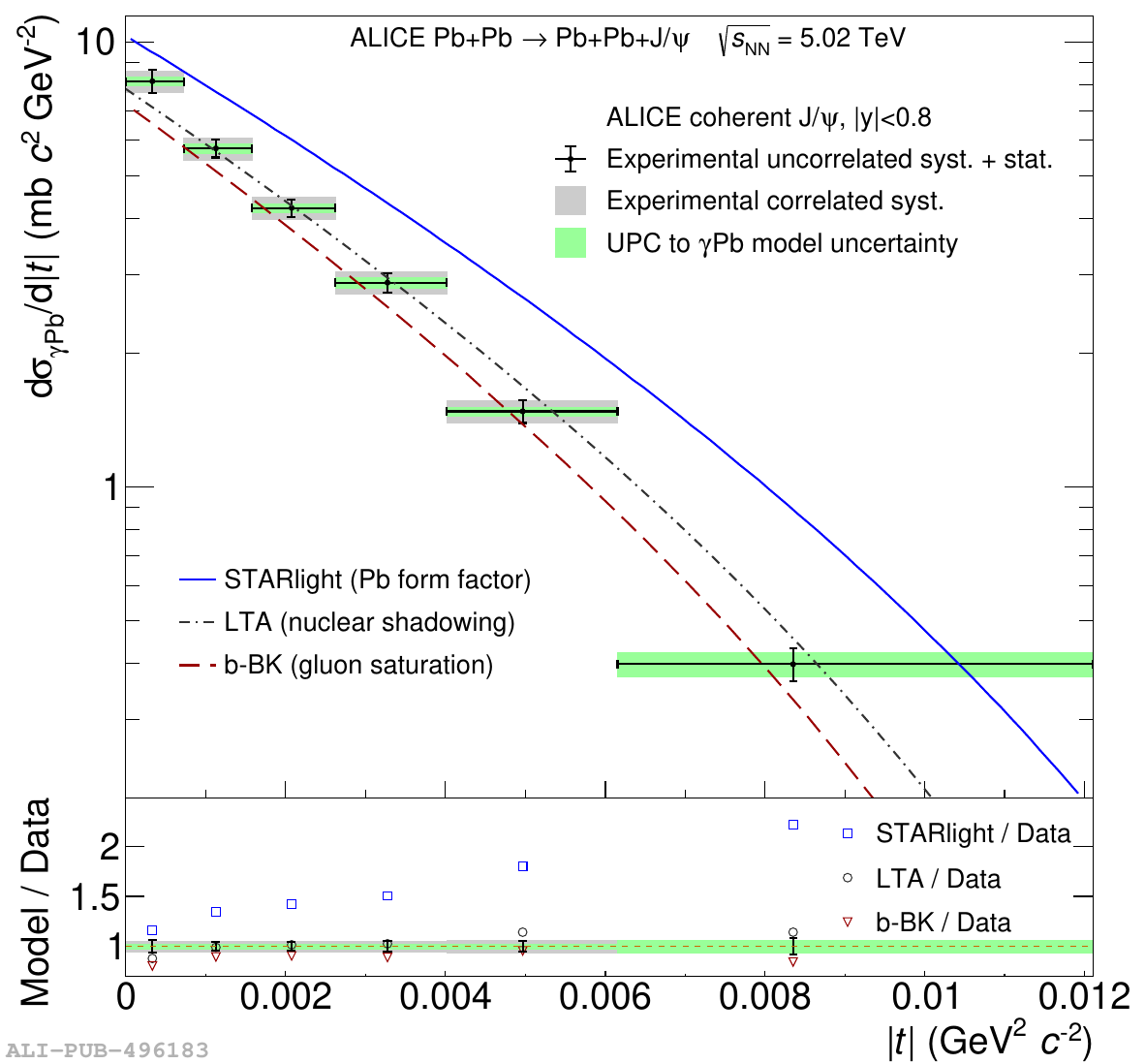}
\includegraphics[width=0.49\textwidth]{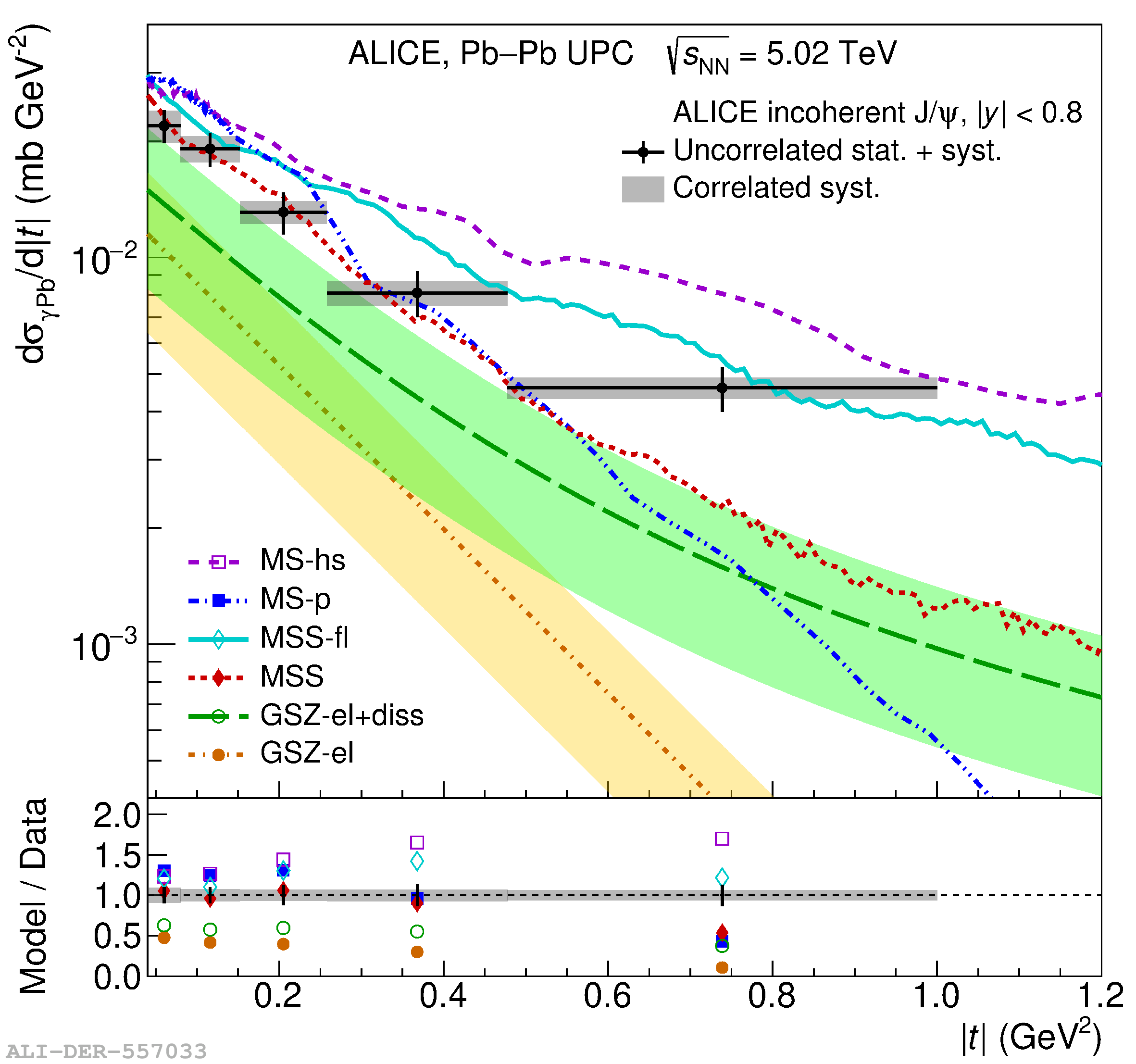}
\caption{The measured dependence of coherent (left) and incoherent (right) \Jpsi photoproduction on the transferred momentum \mant~\protect\cite{ALICE:2021tyx,ALICE:2023gcs} and the comparison with the models (see the text for details). The various uncertainties are depicted as explained in the top panels. The left bottom panel shows the model-to-data ratio in each measured point. The right bottom panel shows the ratio of the integral of the predicted to that of the measured cross section in each interval of \mant. }
\label{fig:tdeps}
\end{figure}

The left panel of Figure~\ref{fig:tdeps} displays the measured \mant-dependence of the coherent cross section and the comparison with three phenomenological models. STARlight~\cite{Klein:2016yzr}, where the \pT spectrum is determined from the nuclear form factor, gives a too high cross section and a slope clearly deviating from the measurement. Dynamic QCD effects are included in the two other models; either through a leading twist approximation of nuclear shadowing~\cite{Guzey:2016qwo}, or via solving the $b$-dependent BK equation, where gluon saturation effects are taken into consideration~\cite{Bendova:2020hbb}. These two models describe the measurement reasonably well. With the new ALICE data from Run~3 and improved detector performance, it should soon be possible to distinguish which of the perturbative QCD calculations provides the better picture. 

The measured evolution of the cross section for incoherent \Jpsi production with \mant is presented in the right panel of Figure~\ref{fig:tdeps} and compared with the predictions of three groups, MS, MSS, and GSZ~\cite{Mantysaari:2017dwh,Mantysaari:2022sux,Guzey:2018tlk}, each offering two scenarios. The slope of the incoherent cross section is sensitive to fluctuations of the transverse profile of the target. The scenarios from the first group (MS-p, MSS, GSZ-el) consider elastic scattering of the photon on a full nucleon; these models generally predict steeper slopes than seen in the data. In the second scenario, subnucleon degrees of freedom are included. MS-hs includes saturation through the IPsat model~\cite{Kowalski:2003hm}, the nucleon is depicted as the sum of three hot spots with variable positions and fluctuations in the saturation scale are considered. The MSS-fl model incorporates perturbative small-$x$ evolution through the solution to the JIMWLK equation. GSZ-el+diss expresses the total incoherent cross section as the sum of the elastic and dissociative parts; both are parameterized based on HERA data and multiplied by a common factor representing nuclear gluon shadowing in the LTA. At high \mant, the data favor the predictions from the second group, where the trend is softened by the inclusion of subnucleon degrees of freedom. Note that the models generally fail to describe the normalization of the cross section, which is mainly linked to scaling from proton to nuclear targets.

\section{Polarization of coherently photoproduced \texorpdfstring{\Jpsi}{J/psi}}
The first measurement of the polarization of coherently photoproduced \Jpsi has been performed by ALICE using Pb--Pb UPCs at $\sqrt{s_{\rm NN}} = 5.02$~TeV~\cite{ALICE:2023svb}. \Jpsi mesons decaying into muon pairs were detected using the forward muon spectrometer and the polarization parameters $\lambda_\theta$, $\lambda_\phi$, and $\lambda_{\theta\phi}$ were extracted from fits to polar and azimuthal angular distributions of \Jpsi yields. The angular distributions were studied in the helicity frame, where the $z$ axis is parallel with the \Jpsi momentum in the Pb--Pb center-of-mass frame. The results are compatible with the transverse polarization, $(\lambda_\theta, \lambda_\phi, \lambda_{\theta\phi}) = (1,0,0)$, and constitute the first experimental evidence for the $s$-channel helicity conservation (SCHC) hypothesis in \Jpsi photoproduction off Pb nuclei. For comparison with previous measurements in electron-proton collisions at HERA, elements of the spin density matrix were extracted; the results were found to be consistent with those from H1 but lower than those from ZEUS, where an electroproduction sample (with higher photon virtuality) was used.

\section{Conclusion and outlook}
This paper presents a summary of recent measurements performed by the ALICE Collaboration based on data from Pb--Pb and p--Pb ultra-peripheral collisions obtained during Run~2 of the LHC, focusing on \Jpsi photoproduction. ALICE has successfully performed comprehensive studies that are crucial to improve the current understanding of the structure of protons and nuclei at high energies. The structure is characterized by the dominance of the gluon component, which is expected to saturate. For nucleons bound in nuclei, the high-energy structure is further modified by the nuclear environment, leading to nuclear shadowing. The new ALICE data provide an important benchmark for refining the predictions of current phenomenological models. The study of the energy dependence of coherent photonuclear production of \Jpsi in Pb--Pb UPCs indicates a strong gluon depletion below $x \sim 10^{-2}$ due to nuclear shadowing. The first analysis of the energy dependence of dissociative photoproduction of \Jpsi off protons at a hadron collider shows agreement with H1 results and represents the first experimental step in the study of fluctuations of the proton substructure. In the first measurements of the dependence of coherent and incoherent \Jpsi photoproduction in Pb--Pb UPCs on Mandelstam $|t|$, the average and variation of the nuclear gluon density have been investigated, respectively. The results of the latter study suggest that quantum fluctuations of the gluon fields need to be taken into account in the available models in order to describe the data. 

In Run~3 (2022--2025) and 4 (2029--2032), ALICE shall be able to perform even more detailed UPC studies and improve the accuracy of the current measurements thanks to the estimated size of data samples~\cite{Citron:2018lsq} and the upgrades on the detectors. Since Run~3, most ALICE detectors operate in continuous readout mode, which reduces the luminosity losses related to the limited efficiency of the trigger systems. ALICE performance in Run~3 particularly benefits from the upgrade to the ITS system; the new design fully relies on monolithic active pixel sensors (MAPS) with a spatial resolution of about 5~\textmu m and significantly reduces the material budget in the vicinity of the interaction point~\cite{ALICE:2013nwm}. The tracking and vertex resolution of the muon spectrometer were also considerably improved thanks to the installation of the Muon Forward Tracker (MFT) based on the same MAPS technology. In Run~4, the scheduled Forward Calorimeter (FoCal) upgrade~\cite{CERN-LHCC-2020-009} will allow to measure gluon distribution functions in the unprecedented Bjorken-$x$ range down to $10^{-6}$~\cite{Bylinkin:2022wkm}.

\printbibliography

\end{document}